\documentclass[twocolumn]{aastex62}

\newcommand{\arcd}{\ifmmode{^{\circle}}\else$'$\fi}
\newcommand{\arcm}{\ifmmode{'}\else$'$\fi}
\newcommand{\arcs}{\ifmmode{''}\else$''$\fi}

\received{May 23, 2019}
\revised{July 22, 2019}
\accepted{July 31, 2019}

\shorttitle{OGLE-UCXB-01}
\shortauthors{Pietrukowicz et al.}

\begin{document}

\title{Discovery of an Outbursting 12.8 Minute Ultracompact X-Ray Binary\footnote{Based
on observations obtained with the 1.3-m Warsaw telescope at the Las Campanas
Observatory of the Carnegie Institution for Science and archival data
from NASA/ESA \textit{Hubble Space Telescope} under program GO 14074
and \textit{Chandra} X-ray Observatory under program GO 17844.}}

\correspondingauthor{Pawel Pietrukowicz}
\email{pietruk@astrouw.edu.pl}

\author[0000-0002-2339-5899]{Pawel Pietrukowicz}
\affil{Warsaw University Observatory \\
Al. Ujazdowskie 4, 00-478 Warszawa, Poland}

\author[0000-0001-7016-1692]{Przemek Mr\'oz}
\affiliation{Warsaw University Observatory \\
Al. Ujazdowskie 4, 00-478 Warszawa, Poland}

\author[0000-0001-5207-5619]{Andrzej Udalski}
\affiliation{Warsaw University Observatory \\
Al. Ujazdowskie 4, 00-478 Warszawa, Poland}

\author[0000-0002-7777-0842]{Igor Soszy\'nski}
\affiliation{Warsaw University Observatory \\
Al. Ujazdowskie 4, 00-478 Warszawa, Poland}

\author[0000-0002-2335-1730]{Jan Skowron}
\affiliation{Warsaw University Observatory \\
Al. Ujazdowskie 4, 00-478 Warszawa, Poland}



\begin{abstract}

We report the discovery of OGLE-UCXB-01, a 12.8 minute variable object
located in the central field of Galactic bulge globular cluster Djorg~2.
The presence of frequent, short-duration brightenings at such an ultrashort
period in long-term OGLE photometry together with the blue color of
the object in \textit{Hubble Space Telescope} images and the detection of
moderately hard X-rays by \textit{Chandra} observatory point to an ultracompact
X-ray binary system. The observed fast period decrease makes the
system a particularly interesting target for gravitational-wave
detectors such as the planned Laser Interferometer Space Antenna.

\end{abstract}



\section{Variable Stars and the OGLE Survey}

Stellar astrophysics owes a lot to observations of variable stars
since they provide valuable information on stellar interiors and
evolution of single and binary systems. Regular photometric observations
of millions of stars help to improve statistics on known variables and
to discover rare objects. The OGLE project is a long-term variability survey
of the Magellanic System and Milky Way stripe visible from Las Campanas
Observatory, Chile. The original science goal of the project in 1992 was
a monitoring of the Galactic bulge in searches for microlensing events.
That area remains the most frequently observed part of the sky by OGLE
with a cadence as short as 20 minutes in the most crowded fields. Since 2010 March
the project is in its fourth phase \citep[OGLE-IV;][]{2015AcA....65....1U}.
OGLE operates the 1.3 m Warsaw telescope equipped with a 32-detector mosaic
camera of a total field of view of 1.4 deg$^2$ and a pixel size of $0\farcs26$.
The previous phase, OGLE-III, was conducted in the years 2001--2009
\citep{2008AcA....58...69U}. A significant fraction of OGLE exposures is
collected in the Cousins $I$ band, while the remaining exposures are taken
in the Johnson $V$ band. Reductions are performed using the difference
image analysis (DIA) technique \citep{1998ApJ...503..325A,2000AcA....50..421W},
developed especially for dense stellar fields.

OGLE has discovered and classified over one million variable stars
of various types, including transient, irregular, and periodic objects
\citep[e.g.,][]{2014AcA....64..177S,2015AcA....65..297S,2015AcA....65..313M}.
Although the sampling of OGLE observations is not adapted for the detection
of millimagnitude variations on time scales of minutes, short-period
variable sources with amplitudes of tenths of a magnitude have been found.
For example, \cite{2017NatAs...1E.166P} announced the discovery of blue
large-amplitude pulsators (BLAPs), hot pulsating stars with periods
in the range 20--40 minutes and optical amplitudes of 0.2--0.4 mag.

\section{Detection of the Unique Variable}

\begin{figure}[b!]
\includegraphics[width=0.48\textwidth]{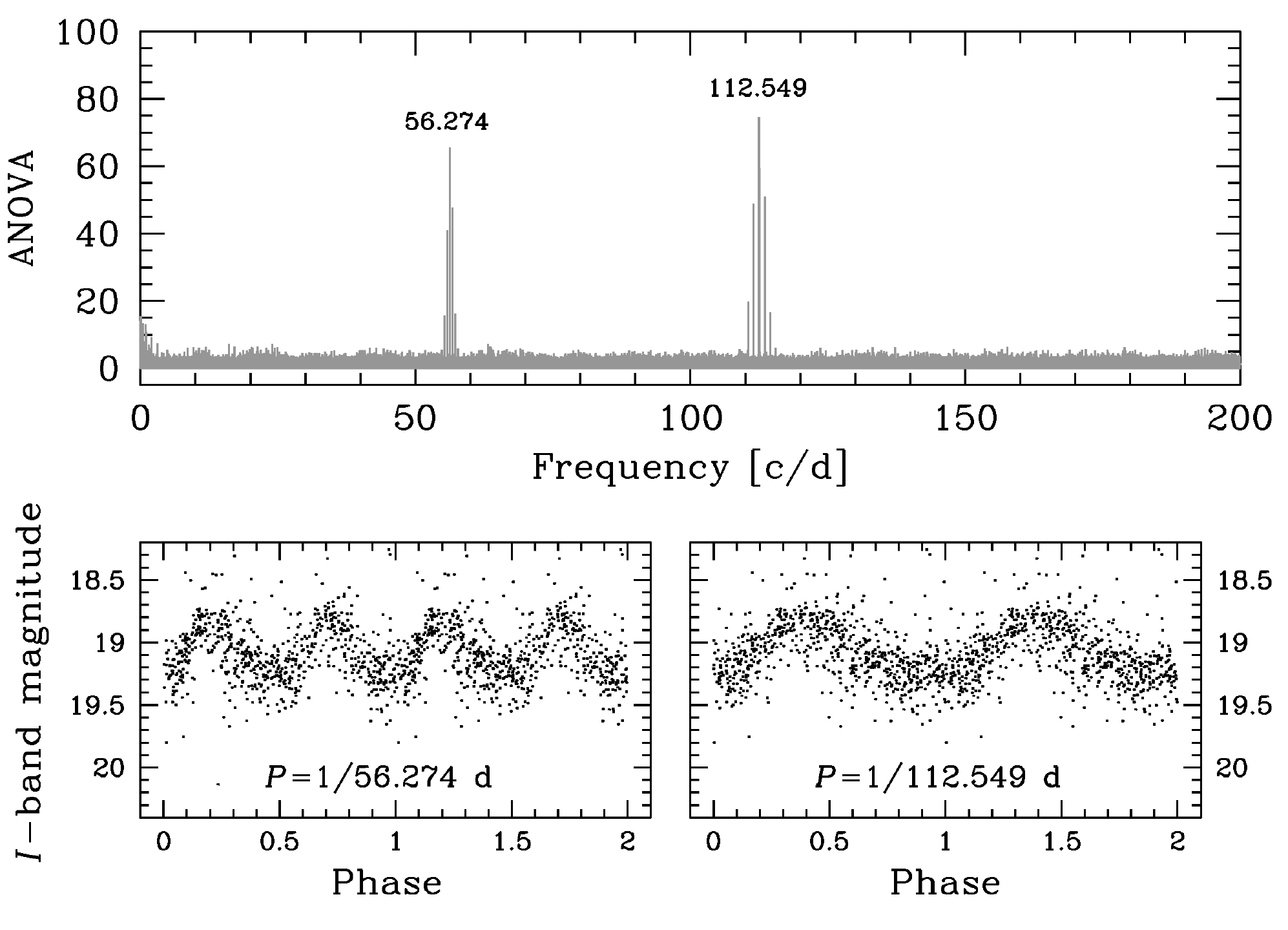}
\caption{ANOVA power spectrum of the discovered variable object
(upper panel). The variability was detected at a frequency of 56.274
cycles per day, corresponding to the period of 25.59 minutes, but the true
modulation is at 112.549 c/d or the period of 12.79 minutes. Lower panels:
phased $I$-band light curve with the two periods. The presented data
come from 2017.}
\label{fig:power}
\end{figure}

In this Letter, we report the discovery of an unusual periodic variable object
with the shortest period ever detected in the OGLE data, that is about 12.79 minutes.
A search for short-period objects was carried out for the Galactic bulge
data covering OGLE-IV seasons 2010--2013. Around 400 million $I$-band
light curves were analyzed in the frequency space from 30 to 100~days$^{-1}$ using the
FNPEAKS\footnote{\tiny http://helas.astro.uni.wroc.pl/deliverables.php?lang=en\&active=fnpeaks}
code. It calculates Fourier amplitude spectra of unequally spaced time-series
data composed of a large number of points ($10^4$ or more). The code substantially
reduces (by a factor of five) the computation time for a discrete Fourier
transform by coadding correctly phased, low-resolution Fourier transforms
of pieces of the large data set interpolated to high resolution. OGLE light
curves with the resulting highest signal-to-noise ratio were subjected to visual
inspection. Among the identified variable stars were BLAPs \citep{2017NatAs...1E.166P}.
Here, a particular attention is paid to the detection BLG511.06.25872 (in the OGLE
database, detection \#25872 in chip 06 of the Galactic bulge field BLG511).
Accurate period determination together with the estimation of its uncertainty
was performed with the TATRY code \citep{1996ApJ...460L.107S}. It
employs periodic orthogonal polynomials to fit the data and the analysis of
variance (ANOVA) statistic to evaluate the quality of the fit. Due to the very
short period of the variable object, all moments were converted from
Heliocentric Julian Date (HJD) to Barycentric Julian Date (BJD$_{\rm TDB}$).
The true period turned out to be even shorter than the minimum searched
period of 14.4 minutes, and the detected signal was an alias (see the power
spectrum and phased OGLE light curves in Fig. \ref{fig:power}).
We notice that the reported period is the only one seen in this star.
Any longer period would be detected in the OGLE data.

\begin{figure}[]
\centering
\includegraphics[width=0.35\textwidth]{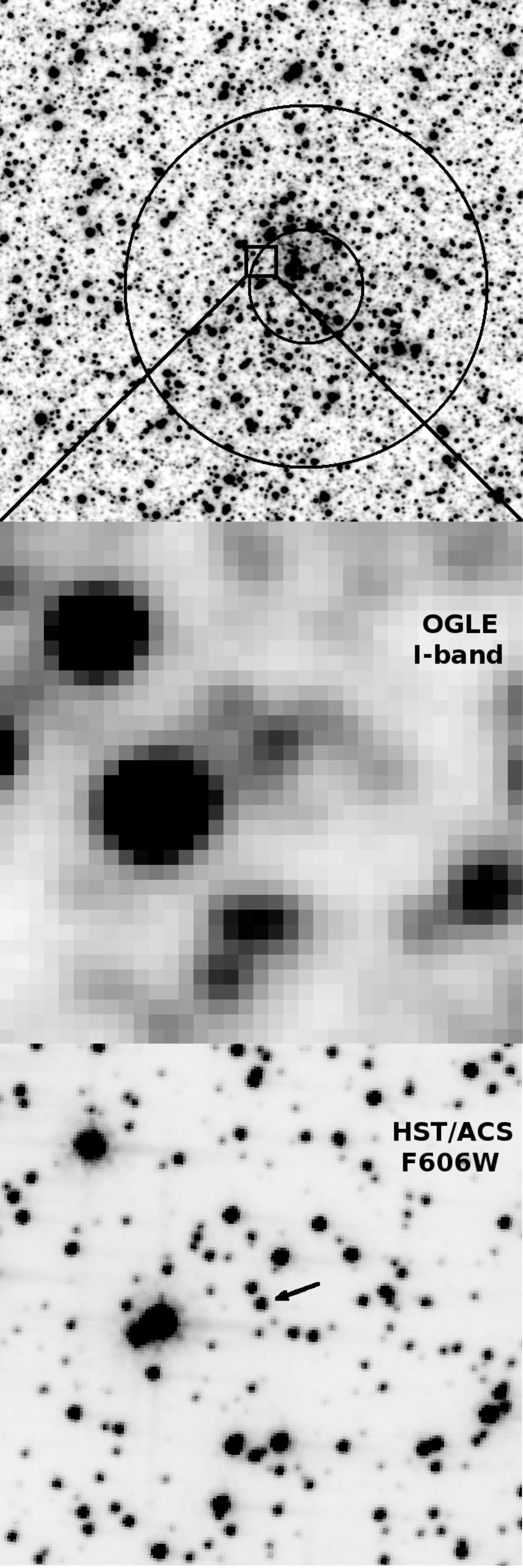}
\caption{Location of the variable object in the field of bulge
globular cluster Djorg~2. Upper panel: 3\arcm$\times$3\arcm~OGLE
$I$-band chart centered on the variable with marked core and half-light
circle of the cluster. North is up and east to the left. Middle
panel: 9\arcs$\times$9\arcs~zoom on the variable that could not be resolved
in the ground-based image. Lower panel: cropped \textit{HST} ACS/WFC image
taken in the F606W filter (broad $V$ band) and covering the same small area.
The variable star is indicated with the arrow. The closest neighbor, located
0\farcs3 northeast of the variable, served as a comparison star.}
\label{fig:chart}
\end{figure}

The variable is located in the field of Galactic bulge globular cluster
Djorg 2 (ESO 456-SC38) at an angular distance of $0\farcm30$ from its center
or at $0.91r_{\rm c}$, where the cluster core radius $r_{\rm c}=0\farcm33$
\citep[from the 2010 version of the catalog by][]{1996AJ....112.1487H}.
However, it was not possible to point the true source in ground-based
frames due to severe blending. Fortunately, the field of Djorg 2 was
imaged with the \textit{Hubble Space Telescope} (\textit{HST}) under program GO 14074.
The cluster was observed with the Wide Field Camera of the Advanced Camera
for Surveys (ACS/WFC) in the optical filter F606W (corresponding
to broad $V$) on 2016 August 19. A week later, on 2016 August 26, it was
observed in near-infrared filters with the Wide Field Camera~3 (WFC3/IR),
F110W (covering $ZY$ bands), and F160W ($H$ band). The \textit{HST} data allowed us
to indicate the true variable star (see Fig. \ref{fig:chart}) at position
$(x,y)=(2368,2588)$ in the ACS/WFC drizzled image jcx324010 or at the
equatorial coordinates $(\alpha,\delta)_{2000.0}=($18:01:50.30, $-$27:49:24.1).

\section{Investigation of the Nature of the New Object}

\begin{figure}[ht!]
\includegraphics[width=0.48\textwidth]{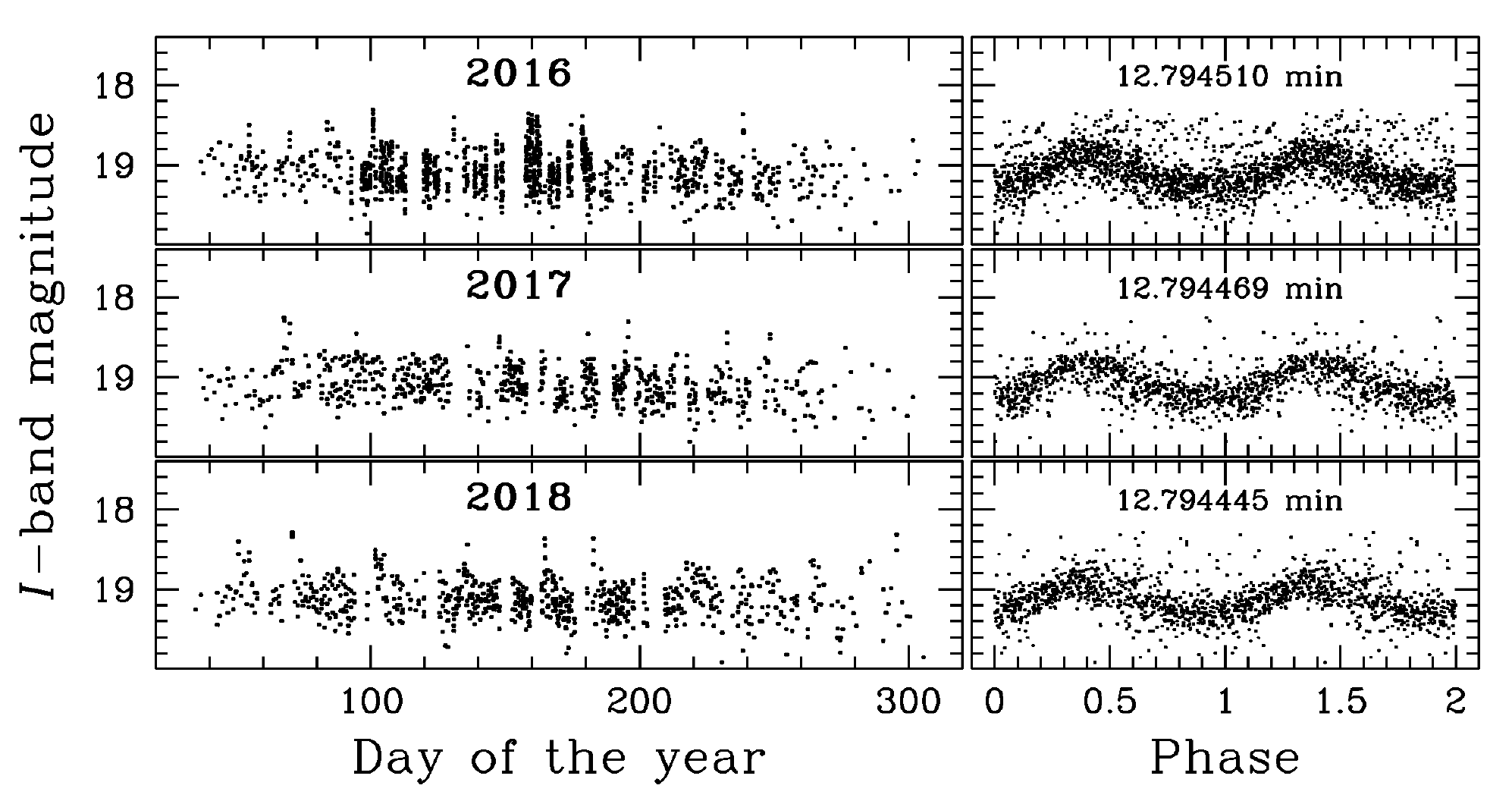}
\caption{OGLE $I$-band light curves of the detected variable from seasons
2016--2018 in the time domain (on the left) and phased with a proper period (on
the right). The presence of outbursts at such short period points to an
ultracompact system. Due to severe blending the real amplitudes of the
outbursts and periodic modulation are expected to be much higher.}
\label{fig:curve}
\end{figure}

\begin{figure}[ht!]
\includegraphics[width=0.48\textwidth]{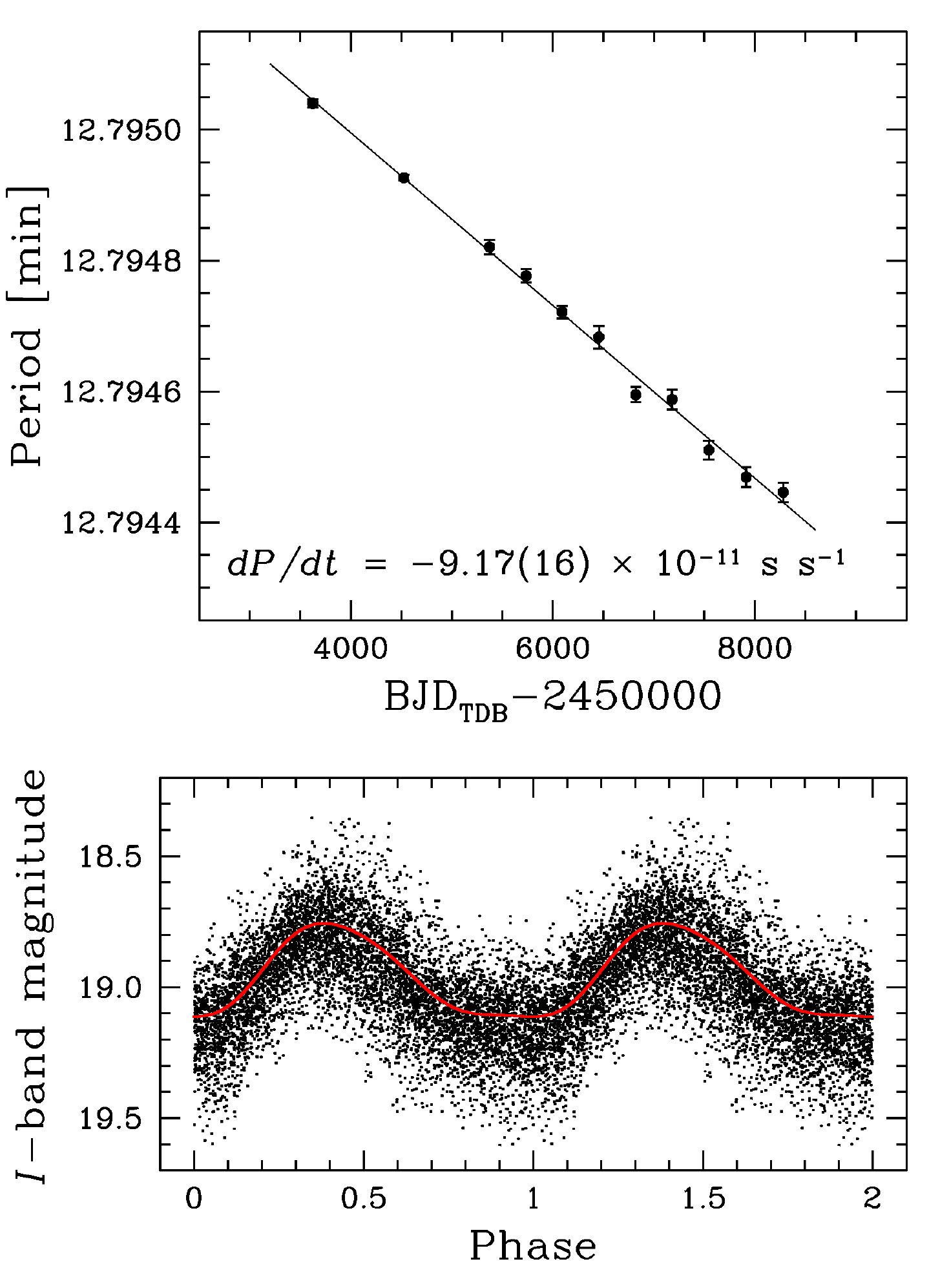}
\caption{Upper panel: constant period decrease observed in the discovered
system. Two points from OGLE-III are averaged over the years 2004--2006 and
2007--2009, while nine points from OGLE-IV are averaged over each year from
2010 to 2018, separately. Lower panel: combination of phased $I$-band light
curves from OGLE-IV seasons 2010--2018 together with the best fit (red line).
Each light curve was cleaned from outbursts and outlying points.}
\label{fig:pdot}
\end{figure}

The object is a periodic variable with an ultrashort period of about
12.79 minutes. Long-term OGLE observations provide some more interesting
information (see Fig. \ref{fig:curve}). In the time domain, we can
notice short brightenings. They last up to several hours and reach
about 1 mag in the $I$ band. Due to severe blending, the true amplitude
of the brightenings is likely much higher, 2 mag or even more.
Another finding is that the period of the variable decreases constantly
with a rate of $-9.16(16) \times 10^{-11}$ s~s$^{-1}$. We phased
light curves from each OGLE-IV season (2010--2018) separately,
cleaned them from outlying points (with $3\sigma$ clipping)
and combined all of them together to one plot (Fig. \ref{fig:pdot}).
Finally, we fit a third-order Fourier series to the data. The combined
$I$-band light curve shows a 0.37 mag bump covering more than half of
the period. The bump is slightly more steep on the raising branch
and the minimum is almost flat. We note that the observed scatter
in the ground-based light curve stems from various reasons: blending,
nonnegligible exposure time in comparison to the variability
period (100~s vs. 767.4~s), and the period decrease.
In Table~\ref{tab:periods}, we compile period values of the variable
star determined from the OGLE photometry over the years 2004--2018.

\begin{table}[]
\centering \caption{Period of the Variable Object in the Years 2004--2018}
\begin{tabular}{lcclr}
\hline
Years  & OGLE  & BJD$_{\rm middle}$ & $P$ [minutes] & $N_{\rm obs}$ \\
       & Phase &           &               & \\
\hline
2004--06 & III & 3624.0977 & 12.795040(6)  &  427 \\
2007--09 & III & 4524.2197 & 12.794926(4)  &  583 \\
2010     & IV  & 5372.5320 & 12.794820(11) &  653 \\
2011     & IV  & 5735.7050 & 12.794776(10) &  818 \\
2012     & IV  & 6092.0027 & 12.794721(10) &  918 \\
2013     & IV  & 6455.5293 & 12.794683(17) &  881 \\
2014     & IV  & 6820.4535 & 12.794594(11) &  845 \\
2015     & IV  & 7180.4325 & 12.794587(15) &  804 \\
2016     & IV  & 7544.2946 & 12.794510(14) & 1132 \\
2017     & IV  & 7913.4080 & 12.794469(15) &  692 \\
2018     & IV  & 8281.0703 & 12.794445(15) &  726 \\
\hline
\end{tabular}
\label{tab:periods}
\medskip
\end{table}

\begin{figure}[ht!]
\includegraphics[width=0.48\textwidth]{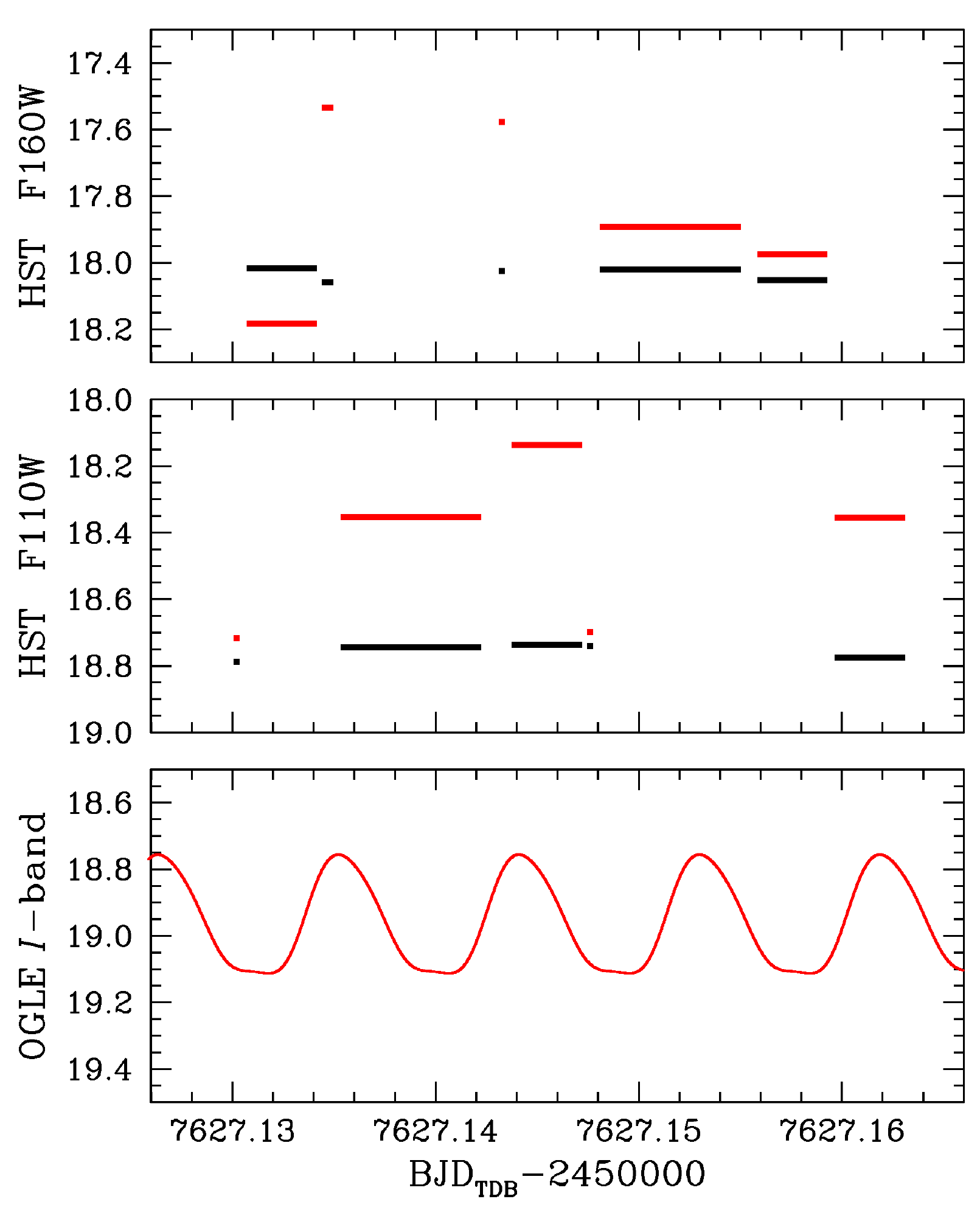}
\caption{Brightness measurements of the discovered object (red lines)
and the comparison star (black lines) obtained from the \textit{HST} WFC3/IR
observations in F160W filter (upper panel) and F110W filter (middle panel)
in comparison with $I$-band variations predicted from the long-term OGLE
data (lower panel). Sections correspond to the start--end moments of the
executed single \textit{HST} exposures. The measured brightness variations
exceed 0.6 mag and are in agreement with the predictions from OGLE.
The nearby comparison star seems to be constant.}
\label{fig:hst}
\end{figure}

Due to significant blending, the true $I$-band amplitude of the
periodic modulation is expected to be much higher. The \textit{HST}
WFC3/IR data show that the amplitude is indeed extreme, although we
cannot determine its exact value. The observations cover over
four variability cycles (Fig. \ref{fig:hst}). Five single images with
various exposure times ranging from about 24 to 599~s were obtained
in each of the two near-infrared filters, F110W and F160W. We determined
magnitudes of the variable star by fitting the point spread function using the
DOLPHOT package \citep{2000PASP..112.1383D}. The WFC3 magnitudes
roughly reflect variations expected from the long-term OGLE observations.
However, one has to remember that the used \textit{HST} filters are centered on
longer wavelengths ($\lambda_{\rm F110W}=1150$ nm, $\lambda_{\rm F160W}=1545$ nm)
in comparison to the OGLE $I$-band filter ($\lambda_I=800$ nm). The
largest difference in brightness reaches 0.58 mag in F110W and 0.65 mag
in F160W, but the true amplitudes are likely slightly higher.

\begin{figure}[ht!]
\includegraphics[width=0.48\textwidth]{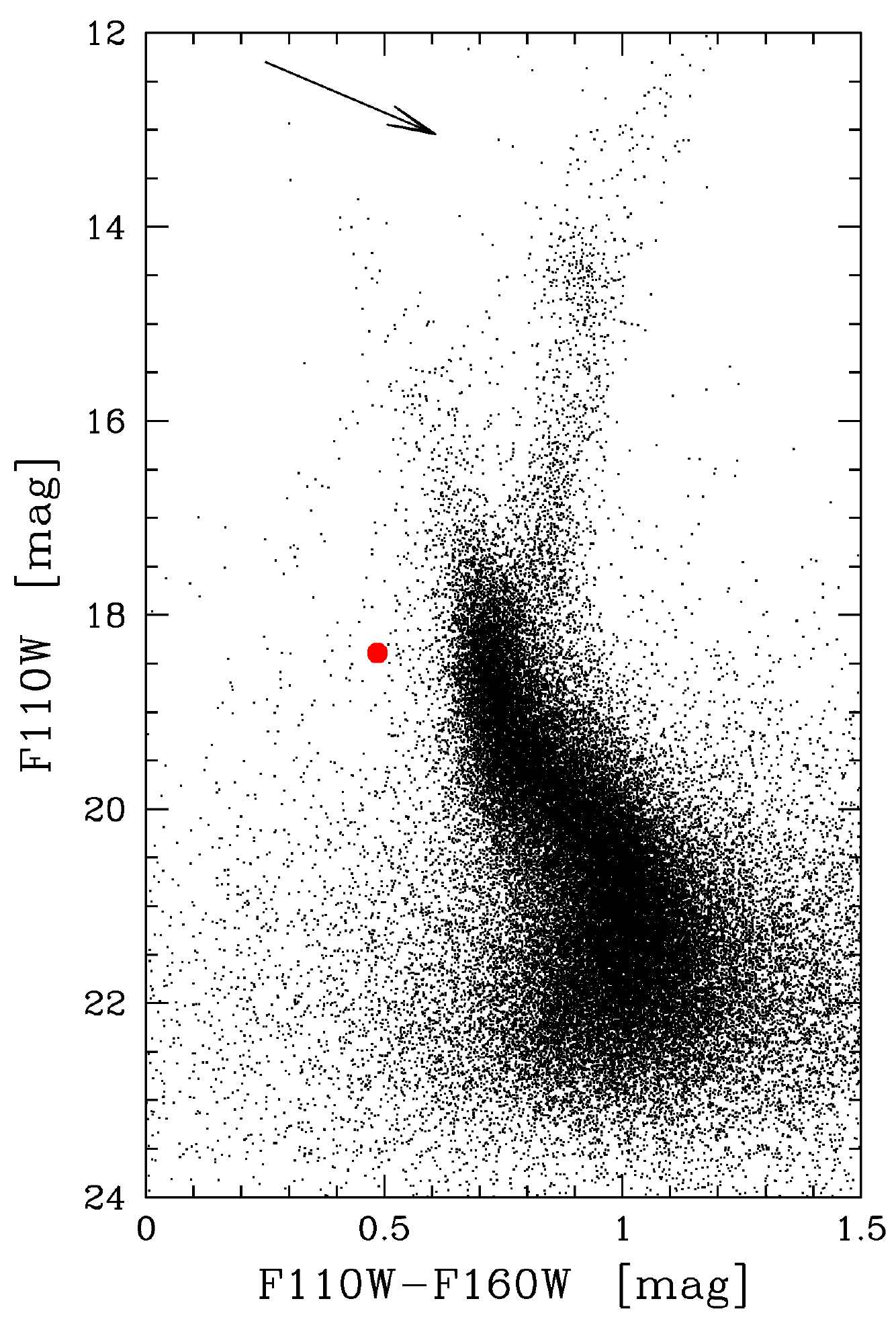}
\caption{Color--magnitude diagram constructed based on \textit{HST} WFC3/IR data for
the cluster field with the marked location of the discovered ultracompact system
(red dot). Its position $\approx0.25$ mag blueward of the main-sequence turnoff
shows that this is a hot object. The arrow represents the reddening vector
in the direction of the variable object as determined from maps in
\cite{2013ApJ...769...88N} and \cite{2015MNRAS.449.1171N}.}
\label{fig:cmd}
\end{figure}

To obtain more information on the nature of the variable object we constructed
a color-magnitude diagram based on the \textit{HST} WFC3/IR data (Fig. \ref{fig:cmd}).
As mean magnitudes of the variable, we adopted values determined from the longest
F110W and F160W exposures (599~s), since these cover a major fraction of the period
($\approx78$\%). In the diagram, the object is located at (F110W$-$F160W, F110W)
= (0.459, 18.352), which is about 0.25 mag blueward of the main-sequence turnoff
point. This position shows that our object is hot. For comparison, the neighboring
constant star located 0\farcs3 northeast of the variable can be found well
within the main sequence at (F110W$-$F160W, F110W) = (0.724, 18.745).

\begin{figure}[ht!]
\includegraphics[width=0.48\textwidth]{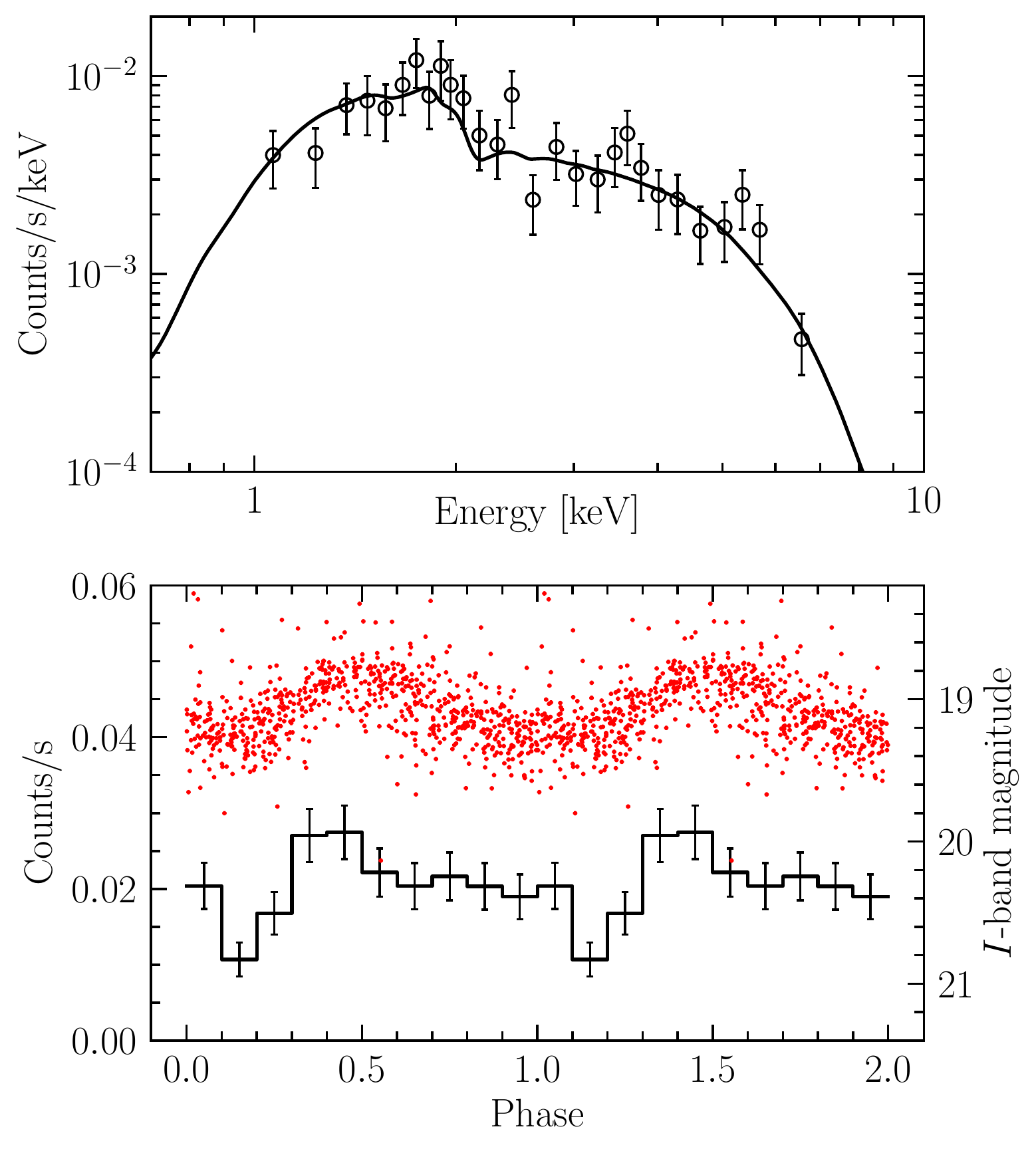}
\caption{Spectral energy distribution (upper panel) and phase distribution
(lower panel) of the X-rays detected with \textit{Chandra} observatory at the
location of the discovered system. The X-ray spectrum can be described by an
absorbed power law with the photon index $\Gamma = 1.22 \pm 0.23$. The X-ray
signal correlates with the optical variability observed by OGLE. The energy
distribution is very similar to the distributions observed in known
ultracompact X-ray binaries. Therefore, we name the new system OGLE-UCXB-01.}
\label{fig:xrays}
\end{figure}

We notice that the investigated system emits X-rays. The area of globular cluster
Djorg~2 was exposed by the \textit{Chandra} satellite observatory for
22.67~ks (about 6.3~hr) on 2017 May 13. There is only one faint
($0.021 \pm 0.001$ cts\,s$^{-1}$) point X-ray source detected within the
cluster area and located $0\farcs64$ from the position of the variable star.
We used CIAO~4.11 software \citep{2006SPIE.6270E..1VF} and CALDB version
4.8.3 to extract the X-ray spectrum of the source, which is shown
in the upper panel of Fig. \ref{fig:xrays}. The spectrum can be well described
by an absorbed power law with the photon index $\Gamma = 1.22 \pm 0.23$.
The estimated equivalent hydrogen column density
($N_{\rm H}=(0.53 \pm 0.20) \times 10^{22}$\,cm$^{-2}$) is consistent with the
interstellar extinction toward the cluster. The X-ray modeling was performed
using the \textsc{Sherpa} package \citep{2001SPIE.4477...76F}.
The total unabsorbed flux in the range 0.5--10\,keV is
$4.8^{+0.5}_{-0.6} \times 10^{-13}$ erg\,s$^{-1}$\,cm$^{-2}$, which
corresponds to the luminosity of $4.4 \pm 0.5 \times 10^{33}$ erg\,s$^{-1}$,
assuming the source is associated with the cluster at the distance of
8.75~kpc \citep{2019A&A...627A.145O}. In the lower panel of Fig. \ref{fig:xrays},
we compare the phased X-ray light curve with the optical light curve from 2017.
We applied a barycentric correction to the data.
Correlation between the X-ray and optical signal is strong with a coefficient
of 0.72, as obtained after binning both phased light curves into 10 bins.

\section{Conclusions}

All facts reported above indicate that the discovered OGLE variable
object is a UCXB. UCXBs are a subgroup of low-mass X-ray binaries
(LMXBs) in which the primary component is either a black
hole or a neutron star. Lack of other periodicities in the long-term
OGLE photometry means that the 12.79 minute signal represents
the orbital period of the binary. Spin period of the primary is likely
too short to be detected in the OGLE data. X-ray emission is an evidence
for accretion processes in the system. The presence of brightenings
lasting several hours and the light curve shape in the optical
regime point to a small accretion disk around the primary.
Our object cannot be a close cataclysmic system of AM CVn type
formed of a white dwarf accretor and a degenerate helium-rich donor.
In such systems, outbursts do not occur at orbital periods shorter than
about 20 minutes \citep{2010PASP..122.1133S,2018A&A...620A.141R}
and the X-ray spectrum is softer, with a peak around or below 1~keV
\citep[e.g.,][]{2004ApJ...614..358S,2006A&A...457..623R}.
We name the newly discovered object OGLE-UCXB-01.

About a third of known UCXBs were found in globular clusters
\citep{2010NewAR..54...87N}. For instance, the prototype of the whole group,
4U 1820-30 or Sgr X-4, resides in Galactic bulge globular cluster NGC 6624
\citep{1987ApJ...315L..49S}. OGLE-UCXB-01 is observed within the core
radius of cluster Djorg~2. It may belong to this cluster. Assuming
the distance to Djorg~2 is 8.75~kpc \citep{2019A&A...627A.145O}
and approximate extinction in this direction $A_V\approx2.4$ mag
\citep[based on the interstellar calculator provided in ][]{2013ApJ...769...88N},
we find that the star with the observed brightness F606W $\approx$ 21.2 mag
(broad $V$ band) would have an absolute brightness $M_V \approx +4.1$ mag,
which is a typical value for UCXBs \citep{2010NewAR..54...87N}.

We could consider the possibility that the new object is a binary
system containing a slowly rotating spinning-up accreting neutron star.
The observed optical modulation would be the spin period of the neutron
star with the measured spin-up rate. In this interpretation, however,
it is difficult to explain the absence of a longer modulation representing
the orbital period of the system. Nevertheless, the newly detected
object requires an optical spectrum that may reveal chemical composition
of the accreted matter. Long-term X-ray monitoring would allow searching
for X-ray bursts and their possible correlation with optical outbursts.
Radial velocity and proper motion measurements should provide the answer
to the question whether the object belongs to globular cluster Djorg 2.
OGLE-UCXB-01 as an ultracompact low-mass binary with a fast period decrease
is expected to be a strong gravitational-wave source in the low-frequency
regime. Once the cluster membership is confirmed or the distance to the system
is well determined, the object may serve as a verification target
for planned space mission LISA.

\acknowledgments

We thank Prof. Tomek Bulik for discussions on the nature of the discovered
system, and Dr. Szymon Kozlowski for help in the reduction of the HST data.
We thank OGLE observers for their contribution to the collection
of the photometric data over the years. The OGLE project has received
funding from the National Science Centre, Poland (grant number
MAESTRO 2014/14/A/ST9/00121 to A.U.). P.M. acknowledges support from the
Foundation for Polish Science (Program START). This Letter makes use of
observations from the NASA/ESA \textit{Hubble Space Telescope}, obtained at
the Space Telescope Science Institute, which is operated by the Association
of Universities for Research in Astronomy, Inc., under NASA contract NAS 5-26555.
The \textit{HST} observations are associated with program GO 14074.
Some scientific results reported in this Letter are based on data
obtained from the \textit{Chandra} Data Archive, under program GO 17844.



\end{document}